%% file: main.tex
\documentclass[sigconf,nonacm]{acmart}

\renewcommand\footnotetextcopyrightpermission[1]{} 

\input{macro}

\makeatletter
\newcommand{\@tightdisplayskips}{%
  \setlength{\abovedisplayskip}{2pt plus 0.5pt minus 0.5pt}%
  \setlength{\belowdisplayskip}{2pt plus 0.5pt minus 0.5pt}%
  \setlength{\abovedisplayshortskip}{1pt plus 0.5pt minus 0.5pt}%
  \setlength{\belowdisplayshortskip}{1pt plus 0.5pt minus 0.5pt}%
}
\g@addto@macro\normalsize{\@tightdisplayskips}
\@tightdisplayskips
\makeatother

\newif\ifshowcomments
\showcommentsfalse

\ifshowcomments
\newcommand{\pc}[1]{\textcolor{red}{\ding{46}~{\sf}~Pingchuan: #1}}
\else
\newcommand{\pc}[1]{}
\fi

\begin{document}

\title{\tool: Safety Rule Evolution for LLM Agents via Inductive Logic Programming}

\author{Pingchuan Ma}
\affiliation{%
  \institution{Zhejiang University of Technology}
  \country{China}}
\email{pma@zjut.edu.cn}

\author{Zhaoyu Wang}
\authornote{Corresponding author.}
\affiliation{%
  \institution{HKUST}
  \country{Hong Kong SAR}}
\email{zwangjz@cse.ust.hk}

\author{Zimo Ji}
\affiliation{%
  \institution{HKUST}
  \country{Hong Kong SAR}}
\email{zjiag@cse.ust.hk}

\author{Yuguang Zhou}
\affiliation{%
  \institution{HKUST}
  \country{Hong Kong SAR}}
\email{ygzhou02@gmail.com}

\author{Zhantong Xue}
\affiliation{%
  \institution{HKUST}
  \country{Hong Kong SAR}}
\email{zxueai@connect.ust.hk}

\author{Zongjie Li}
\affiliation{%
  \institution{HKUST}
  \country{Hong Kong SAR}}
\email{zligo@cse.ust.hk}

\author{Shuai Wang}
\affiliation{%
  \institution{HKUST}
  \country{Hong Kong SAR}}
\email{shuaiw@cse.ust.hk}

\author{Xiaoqin Zhang}
\affiliation{%
  \institution{Zhejiang University of Technology}
  \country{China}}
\email{zhangxiaoqinnan@gmail.com}

\renewcommand{\shortauthors}{Ma et al.}

\begin{abstract}
Large language model (LLM) agents increasingly automate complex tasks by
integrating language models with external tools and environments. However,
their autonomy poses significant safety risks: agents may execute
destructive commands, leak sensitive data, or violate domain constraints.
Existing safety approaches face a fundamental tradeoff: hand-crafted rules
are interpretable but brittle, with overly conservative rules blocking safe
operations (high false positives) while permissive rules miss unsafe
behaviors (high false negatives). Neural classifiers lack the
interpretability required for safety-critical deployments.

We present \tool, a framework that automatically evolves deployed
expert-designed safety rules from user safe/unsafe annotations through
counterexample-guided inductive synthesis (CEGIS) guided by inductive logic
programming (ILP). Starting from the expert rules and a stream of annotated
traces, \tool iteratively evaluates rules, mines false-positive and
false-negative counterexamples, uses ILP to learn which predicates
discriminate them, generates candidate rule edits, and verifies candidates
to select the best revision. The key insight is that ILP efficiently
identifies predicates that appear frequently in false negatives but rarely
in false positives (or vice versa), dramatically pruning the exponential
search space of rule edits. This continues until convergence, producing
interpretable rules that balance precision and recall.

We evaluate \tool on 291 execution traces spanning code execution and
embodied agent domains. \tool raises rule F1 to 0.98 and 0.93 across the two
domains, achieving up to 94\% false positive reduction while maintaining high
recall, and converges within 4--5 iterations. The ILP-guided approach
achieves up to 4.8$\times$ higher F1 than heuristic CEGIS. The learned rules
are human-readable, auditable, and generalize to unseen scenarios.
\end{abstract}
\maketitle

\input{introduction}

\input{background}

\input{approach}

\input{setup}

\input{evaluation}

\input{discussion}

\input{related}

\section{Conclusion}

We present \tool, the first tool that \emph{evolves} an LLM agent's deployed
safety rules rather than authoring them from scratch. Its key idea is that the
safe/unsafe annotations an agent accrues in operation are counterexamples, making
rule maintenance a natural ILP-guided CEGIS problem: counterexamples localize
failures, ILP finds the few predicates that separate them, and constrained
operators turn those predicates into auditable edits. Across code-execution and
embodied-agent traces, \tool raises rule F1 to 0.98 and 0.93 within a few
iterations, and practitioners rate the resulting rules far more interpretable and
trustworthy than a neural classifier trained on the same annotations---giving the
symbolic guardrail layer the adaptivity of a learned classifier while keeping it
deployable.

\bibliographystyle{ACM-Reference-Format}
\bibliography{bib/ref}

\end{document}

%% file: macro.tex
\usepackage{graphicx}
\usepackage[many]{tcolorbox}
\usepackage{mathtools}
\usepackage{booktabs}
\usepackage{xcolor}
\usepackage{threeparttable}
\usepackage{multirow}
\usepackage{tikz}
\usetikzlibrary{positioning,calc,fit,arrows.meta}
\usepackage[ruled,linesnumbered]{algorithm2e}
\usepackage{url}

\usepackage{syntax}
\usepackage{framed}
\usepackage{listings}
\usepackage{moresize}
\usepackage{wrapfig}
\usepackage{xspace}
\usepackage{enumitem}
\usepackage{pifont}
\usepackage{makecell}
\usepackage{subcaption}

\newcommand{\cmark}{\ding{51}}%
\usepackage{thmtools}
\declaretheoremstyle[spaceabove=\topsep,notefont=\normalfont\itshape]{mystyle}

\newcommand{\F}{Fig.}

\newcommand{\T}{Table}
\newcommand{\A}{Alg.}
\newcommand{\Def}{Def.}

\newcommand{\ignore}[1]{}

\newcommand{\tool}{\textsc{AutoSpec}\xspace}

\newcommand{\parh}[1]{\noindent\textbf{#1}}

\definecolor{lightgray}{rgb}{.9,.9,.9}
\definecolor{darkgray}{rgb}{.4,.4,.4}
\definecolor{commentgreen}{RGB}{63,127,95}
\definecolor{myGreen}{RGB}{61, 161, 61}
\definecolor{myBrown}{RGB}{196, 89, 17}
\definecolor{pptgreen}{RGB}{84,130,53}
\definecolor{pptred}{RGB}{176,35,24}

\colorlet{myPurple}{blue!40!red}

\SetCommentSty{tcpblue}

\lstdefinelanguage{AgentSpec}{
  keywords={rule, trigger, check, enforce, end, stop, user_inspection, llm_self_reflect, none, invoke_action, true, false},
  keywordstyle=\color{violet}\bfseries,
  ndkeywords={},
  ndkeywordstyle=\color{darkgray}\bfseries,
  identifierstyle=\color{black},
  sensitive=true,
  comment=[l]{\%},
  morecomment=[s]{/*}{*/},
  commentstyle=\color{commentgreen}\ttfamily,
  stringstyle=\color{red}\ttfamily,
  morestring=[b]',
  morestring=[b]"
}

\usepackage{microtype}


%% file: introduction.tex
\section{Introduction}
\label{sec:intro}

Large language model (LLM) agents~\cite{yao2023react,schick2023toolformer,xi2025rise} have emerged as a transformative paradigm for automating complex tasks. By integrating LLMs with external tools, memory, and planning capabilities~\cite{mohammadi2025evaluation}, these agents can execute code, manipulate files, control robots, and interact with physical environments. This has enabled applications ranging from software development assistants~\cite{qian2024chatdev,hong2024metagpt} to autonomous driving~\cite{wen2024dilu} and household robotics~\cite{ahn2022can}. However, the very autonomy that makes LLM agents powerful also makes them dangerous: an agent tasked with ``cleaning up temporary files'' may execute ``\texttt{rm -rf /}'' and a household robot may place a laptop near a running stove~\cite{yuan2024r}.

\parh{The Safety Challenge.} Ensuring LLM agent safety is fundamentally
challenging. Unlike traditional software, LLM agents operate in open-ended
environments with natural language inputs, making pre-deployment verification
intractable. As a result, even well-tested agents can exhibit novel failure
modes in deployment~\cite{zhang2025which}.

\parh{Key Motivation: Expert Rules Must Evolve with the Agent.} In practice,
deployed agents rarely start from a blank slate: they ship with a set of
expert-designed safety rules that catch known hazards, such as blocking
\texttt{rm -rf /} or restricting access to sensitive
files~\cite{wang2025agentspec}. These hand-crafted rules encode valuable domain
expertise and are interpretable and auditable, which is why practitioners rely on
them. The difficulty is that the rules are static while the agent is not: models
are updated, new tools are integrated, and prompts are revised, continually
shifting the boundary between safe and unsafe behavior. Risks also frequently
emerge from \emph{combinations} of actions rather than individual ones (reading
credentials is benign and calling an external API is routine, but doing one after
the other may constitute exfiltration), which no fixed rule set anticipates in
advance. Consequently, rules that were well-calibrated at deployment gradually
drift, raising false alarms on benign workflows or missing newly introduced
hazards, and maintaining them by hand demands scarce expert attention that cannot
keep pace with how fast agent environments change.

Crucially, running agents generate exactly the signal needed to keep these rules
current. As agents execute, users and reviewers routinely annotate individual
actions as safe or unsafe, e.g., by approving or rejecting the borderline tool
calls that agents surface for confirmation. Together with the structured
tool-call traces that agent frameworks already record, these annotations
accumulate into a stream of labeled execution traces. Our key observation is that
this feedback can be used to \emph{automatically refine the initial
expert-designed rules}, adapting them to the evolving agent environment rather
than discarding the expertise they encode or rewriting them from scratch. This
reframes safety-rule maintenance from a manual, one-off authoring task into a
continuous, annotation-driven \emph{evolution} problem, which is the problem
\tool addresses.

Current approaches to agent safety fall into two categories, each with critical
limitations. \emph{Static rule-based approaches}~\cite{wang2025agentspec} define
safety constraints as hand-crafted rules (e.g., ``block any command containing
\texttt{rm -rf /}''). While interpretable and auditable, these rules face a
fundamental precision-recall tradeoff: overly conservative rules flag benign
operations as unsafe (high false positives), disrupting legitimate workflows,
while permissive rules miss novel unsafe patterns (high false negatives),
leaving vulnerabilities unaddressed. Manually balancing this tradeoff requires
deep domain expertise and continuous maintenance as agent behaviors evolve.
\emph{Neural classification approaches}~\cite{inan2023llama} train classifiers
to distinguish safe from unsafe behaviors. While more adaptive, these black-box
models cannot explain \emph{why} a behavior is flagged, making them unsuitable
for safety-critical deployments where auditability and trust are paramount.
Critically, neither category turns the accumulating safe/unsafe annotations into
automatic updates of the deployed expert rules: static rules must be revised by
hand, while neural classifiers \emph{replace} the interpretable rules rather than
\emph{refining} them. What is missing is a way to evolve the existing symbolic
rules directly from this feedback while preserving their interpretability.

\parh{Key Insight: Rule Evolution as CEGIS.} We observe that this evolution problem, starting from the deployed expert rules and systematically refining them against the annotated traces, has a natural formulation as \emph{counterexample-guided inductive synthesis} (CEGIS)~\cite{solar2006combinatorial,jha2010oracle}. Here the human annotations play the role of counterexamples: an action labeled unsafe that the current rules let through, or one labeled safe that they block, is exactly an instance where the rules disagree with ground truth. In CEGIS, a synthesizer proposes candidate programs, a verifier checks them against specifications, and counterexamples from failed verification guide the next synthesis iteration. Applied to safety rules, this means: (1) evaluate the current rules on the annotated execution traces, (2) mine false positives (safe traces incorrectly flagged) and false negatives (unsafe traces missed), (3) synthesize rule edits that resolve these counterexamples, and (4) verify candidates to select the best revision. This loop guarantees progress: each counterexample eliminates at least one incorrect candidate.

However, naive CEGIS faces a critical challenge: the search space of possible rule edits is exponential in the number of available predicates (e.g., \texttt{is\_system\_path}, \texttt{writes\_to\_disk}, \texttt{near\_danger\_zone}). Enumerating all combinations of predicates and edit operations (add conjunct, add exception, relax, create disjunctive branch) quickly becomes intractable. Prior work on CEGIS relies on heuristic search strategies~\cite{pu2018selecting} that explore this space uniformly, leading to slow convergence and suboptimal rules.

\parh{Our Approach: ILP-Guided CEGIS.} We propose \tool, a framework that
addresses this challenge by using \emph{Inductive Logic Programming}
(ILP)~\cite{muggleton1994inductive,law2020ilasp} to guide the CEGIS search. The
key insight is that ILP can efficiently learn which predicates are
\emph{discriminating} for counterexamples, namely predicates that appear
frequently in false negatives but rarely in false positives (or vice versa). By
prioritizing these predicates, \tool prunes the search space dramatically,
focusing on edits most likely to resolve counterexamples while maintaining rule
interpretability.

Specifically, given false-positive and false-negative counterexamples, \tool formulates an ILP learning task where the goal is to learn a hypothesis that distinguishes unsafe traces (positive examples) from safe traces (negative examples). The ILP solver (ILASP~\cite{law2020ilasp}) outputs a ranked list of predicates with suggested edit operations (e.g., ``add \texttt{is\_temp\_file} as exception to reduce false positives''). \tool then generates candidate rule edits guided by these suggestions, verifies each candidate on the full dataset, and selects the best revision using a score that prioritizes F1 while also favoring precision, fewer unresolved counterexamples, and simpler rules. This process iterates until convergence or a target score threshold is reached.

\parh{Contributions.} Our contributions are threefold:

\begin{enumerate}[leftmargin=*,itemsep=2pt]
\item \emph{Conceptually}, we frame safety-rule maintenance as evolving deployed expert rules from user safe/unsafe annotations, and cast it as ILP-guided CEGIS in which annotations act as counterexamples and ILP prunes the exponential space of rule edits.

\item \emph{Technically}, we build \tool, the first safety-evolution tool: it refines symbolic guardrails via counterexample mining, ILP-guided predicate ranking, candidate generation, and verification, producing auditable rules that complement existing defenses and deploy directly.

\item \emph{Empirically}, across 291 traces from code-execution and embodied-agent domains, \tool reaches F1 of 0.98 and 0.93, with ILP guidance yielding up to 4.8$\times$ improvement over heuristic CEGIS; the rules converge in 4--5 iterations, generalize to unseen tasks, and are rated far more interpretable and trustworthy than neural classifiers by practitioners.
\end{enumerate}

%% file: background.tex
\section{Background}
\label{sec:background}

\subsection{LLM Agent Architecture}

An LLM agent consists of a language model $\mathcal{M}$, a set of tools $\mathcal{T} = \{t_1, t_2, \ldots, t_k\}$, and an execution loop that iteratively selects and invokes tools based on the model's output. At each step $i$, the agent observes the current state $s_i$ (including conversation history, tool outputs, and environment state), generates an action $a_i = (t, \mathit{args})$ specifying a tool $t \in \mathcal{T}$ and its arguments, and receives an observation $o_i$ from the environment.

While this architecture enables powerful autonomous behavior, it also introduces significant safety risks. The agent's autonomy means it can execute actions with real-world consequences, such as deleting files, sending network requests, or controlling physical actuators, based solely on the LLM's predictions. Since LLMs are stochastic and can produce unexpected outputs, especially when faced with adversarial inputs~\cite{perez2022red, liu2024exploring} or distribution shift~\cite{ren2025llms}, agents may take unsafe actions even when given benign user requests. These risks are exacerbated in production deployments where agents operate with elevated privileges and limited human oversight.

To formalize agent behavior for safety analysis, we represent each agent session as an execution trace.

\begin{definition}[Execution Trace]
An execution trace $\tau = \langle e_1, e_2, \ldots, e_n \rangle$ is a sequence
of events, where each event $e_i = (t_i, \mathit{type}_i, s_i, a_i, o_i)$
records the timestamp $t_i$, event type $\mathit{type}_i \in
\{\texttt{before\_action}, \allowbreak  \texttt{after\_action}, \texttt{state\_change}\}$,
state snapshot $s_i$, action $a_i$, and tool output $o_i$.
\end{definition}

Execution traces capture the complete history of an agent's interaction with its environment, enabling post-hoc analysis of agent behavior. Each trace is labeled as \emph{safe} or \emph{unsafe} based on whether the agent's actions violated safety properties. For example, a trace where the agent executes \texttt{rm -rf /home/user/.ssh/} would be labeled unsafe, while a trace that only reads files would be labeled safe. These labeled traces form the training data $\mathcal{D}_L$ for rule refinement (see \S\ref{sec:overview}).

\subsection{Predicate-Based Safety Constraints}

We model safety knowledge as a set of predicate-based constraints over agent traces. The central object is the predicate library $\mathcal{P}$: each predicate captures an interpretable property of an event or its local context.  Constraints (or \emph{rules}) are constructed by composing these predicates into conjunctions, exceptions, and disjunctive branches. In our implementation, these constraints are encoded in the AgentSpec framework~\cite{wang2025agentspec}, but the formulation used in this paper depends only on predicate evaluation over traces rather than on any particular DSL.

\parh{Predicate Library.} \tool maintains a cross-domain predicate library $\mathcal{P}$ organized by application domain. Each predicate is a boolean function $p: \mathcal{E} \rightarrow \{\top, \bot\}$ (true or false) that evaluates properties of an event, implemented as a program function that asserts whether the predicate holds on a specific agent event. \T~\ref{tab:predicates} shows representative predicates from each domain. For example, \texttt{destructive\_os\_inst} checks if a code execution action invokes destructive operations like \texttt{rm -rf} or \texttt{os.remove}, while \texttt{near\_danger\_zone} checks if a robot is approaching hazardous areas (stove, knife, fire). Our evaluation uses domain-specific predicate libraries ranging in size from 16 to 24 predicates per domain; the full lists are provided in the supplementary material.

\begin{table}[t]
\centering
\caption{Example predicates from the predicate library $\mathcal{P}$.}
\label{tab:predicates}
\resizebox{0.9\columnwidth}{!}{%
\begin{tabular}{@{}ll@{}}
\toprule
\textbf{Domain} & \textbf{Example Predicates} \\
\midrule
Code & \texttt{destructive\_os\_inst}, \texttt{involve\_system\_file}, \\
     & \texttt{high\_risk\_syscall}, \texttt{execute\_script}, \\
     & \texttt{write\_to\_io}, \texttt{submit\_post\_request} \\
\midrule
Embodied & \texttt{near\_danger\_zone}, \texttt{handling\_fragile\_object}, \\
         & \texttt{carrying\_liquid}, \texttt{source\_is\_heat} \\
\bottomrule
\end{tabular}%
}
\end{table}

\parh{Safety Constraints.} Building upon the foundational predicates in $\mathcal{P}$, we construct safety constraints that operate fundamentally at the granularity of individual events. Formally, we define a rule $r$ as a logical formula composed of these base predicates, and establish its activation as follows:

\begin{definition}[Rule Firing]
\label{def:rule-firing}
Given a trace $\tau$ and a safety constraint $r$, rule $r$ fires on event $e_i \in \tau$ if the predicate condition associated with $r$ evaluates to $\top$ on $e_i$.
\end{definition}

\begin{example}
For instance, a representative safety rule $r \in \mathcal{R}$ is:
$$r(e) = \texttt{destructive\_os\_inst}(e) \land \neg \texttt{is\_temp\_file}(e)$$
which evaluates to $\top$ (fires) to block destructive file operations unless they target temporary files. 
\end{example}

While rules like this evaluate at the localized event level, overall trace safety is determined by aggregating these individual firings:

\begin{definition}[Safety Classification]
\label{def:safety-classification}
A trace $\tau$ is classified as \emph{unsafe} by a rule set $\mathcal{R}$ if there exists a rule $r \in \mathcal{R}$ that fires on an event $e_i \in \tau$. Otherwise, $\tau$ is classified as \emph{safe}.
\end{definition}

These two definitions establish a deliberate two-level structure: rules operate on individual events (\Def~\ref{def:rule-firing}), while supervision is provided at trace granularity (\Def~\ref{def:safety-classification}). This reflects common practice where a human operator observes that an agent session produced an undesirable outcome without necessarily identifying the exact offending action. \tool bridges these two levels in its ILP encoding by recording which predicates hold at each event within a trace, then learning trace-level classification rules over these per-event facts (\S\ref{sec:ilp-guided-predicate-learning}).

\subsection{Inductive Logic Programming and CEGIS}

\tool combines two synthesis paradigms: inductive logic programming (ILP) for learning discriminating predicates, and counterexample-guided inductive synthesis (CEGIS) for iterative refinement. This subsection provides the formal background for both.

\parh{Inductive Logic Programming.} ILP~\cite{muggleton1994inductive,law2020ilasp} is a machine learning approach that learns logic programs from examples. Given a background knowledge base $B$, positive examples $E^+$, and negative examples $E^-$, an ILP solver searches for a hypothesis $H$ (a set of Horn clauses) such that:
\begin{align*}
B \cup H &\models e^+ \quad \forall e^+ \in E^+ \quad \text{(completeness)} \\
B \cup H &\not\models e^- \quad \forall e^- \in E^- \quad \text{(consistency)}
\end{align*}

The hypothesis space is typically constrained by mode declarations that specify the structure of learnable clauses. For instance, a mode declaration might restrict hypotheses to conjunctions of predicates from a fixed vocabulary, with each predicate appearing at most once. Modern ILP solvers like ILASP~\cite{law2020ilasp} support answer set programming semantics and can optimize for hypothesis simplicity through length-based scoring.

\parh{Counterexample-Guided Inductive Synthesis.} CEGIS~\cite{solar2006combinatorial} is an iterative synthesis framework with two components: a synthesizer that proposes candidate solutions, and a verifier that checks candidates against a specification. The loop proceeds as follows: \ding{202}~the synthesizer generates a candidate $c$ from the current constraint set; \ding{203}~the verifier checks whether $c$ satisfies the specification on all inputs; \ding{204}~if verification succeeds, it returns $c$, otherwise it produces a counterexample $x$ where $c$ fails; and \ding{205}~$x$ is added to the constraint set and the loop repeats.

Formally, let $\mathcal{C}$ be the set of constraints (initially empty), $\phi$ the specification, and $\mathcal{H}$ the hypothesis space. At iteration $i$, the synthesizer solves:
\[
c_i = \arg\max_{h \in \mathcal{H}} \text{score}(h) \quad \text{s.t.} \quad h \text{ satisfies } \mathcal{C}_i
\]
The verifier then checks $c_i \models \phi$. If not, it finds a counterexample $x$ such that $c_i(x) \neq \phi(x)$, and updates $\mathcal{C}_{i+1} = \mathcal{C}_i \cup \{x\}$.

\tool instantiates this framework for safety rule refinement: the synthesizer uses ILP to propose predicate-based rule edits, and the verifier evaluates candidates on labeled execution traces to identify false positives and false negatives as counterexamples.

%% file: approach.tex
\section{Overview}
\label{sec:overview}

The central challenge is that safety-rule refinement lives in a large combinatorial space while the output must remain interpretable as a symbolic safety constraint. \tool addresses this by combining the feedback loop of CEGIS with ILP-guided predicate selection: counterexamples indicate where rules fail, ILP identifies promising predicates for repair, and constrained edit operators translate these signals into verifiable candidate rules. The remainder of this section introduces the optimization problem (\S~\ref{sec:problem-formulation}), summarizes the workflow (\S~\ref{sec:workflow}), and situates \tool within the broader defense-in-depth stack (\S~\ref{sec:application-scope}).

\subsection{Problem Formulation}
\label{sec:problem-formulation}
To systematically navigate this refinement process, we frame rule evolution as a discrete optimization problem. The goal is to maximize the performance of an initial rule set over a labeled dataset by applying a sequence of logical edits.

\begin{definition}[Rule Evolution Problem]
Given an initial rule set $\mathcal{R}_0$, a set of labeled traces $\mathcal{D} = \{(\tau_i, y_i)\}_{i=1}^{N}$ where $y_i \in \{\texttt{safe}, \texttt{unsafe}\}$, and a predicate library $\mathcal{P}$, find a rule set $\mathcal{R}^*$ that maximizes the validation score on $\mathcal{D}$:
\[
\mathcal{R}^* = \arg\max_{\mathcal{R} \in \mathit{Edit}^*(\mathcal{R}_0, \mathcal{P})} \; S(\mathcal{R})
\]
where $\mathit{Edit}^*(\mathcal{R}_0, \mathcal{P})$ is the set of all rule sets reachable from $\mathcal{R}_0$ through a sequence of edit operations using predicates from $\mathcal{P}$, and $S(\cdot)$ is the candidate score defined later in this section.
\end{definition}

Because the search space $\mathit{Edit}^*(\mathcal{R}_0, \mathcal{P})$ grows exponentially with the number of available predicates and edit steps, exhaustive enumeration is computationally intractable. To overcome this bottleneck, \tool utilizes an ILP-guided CEGIS approach to efficiently navigate this space and discover optimal rules.

\begin{figure}[t]
    \centering
    \includegraphics[width=\columnwidth]{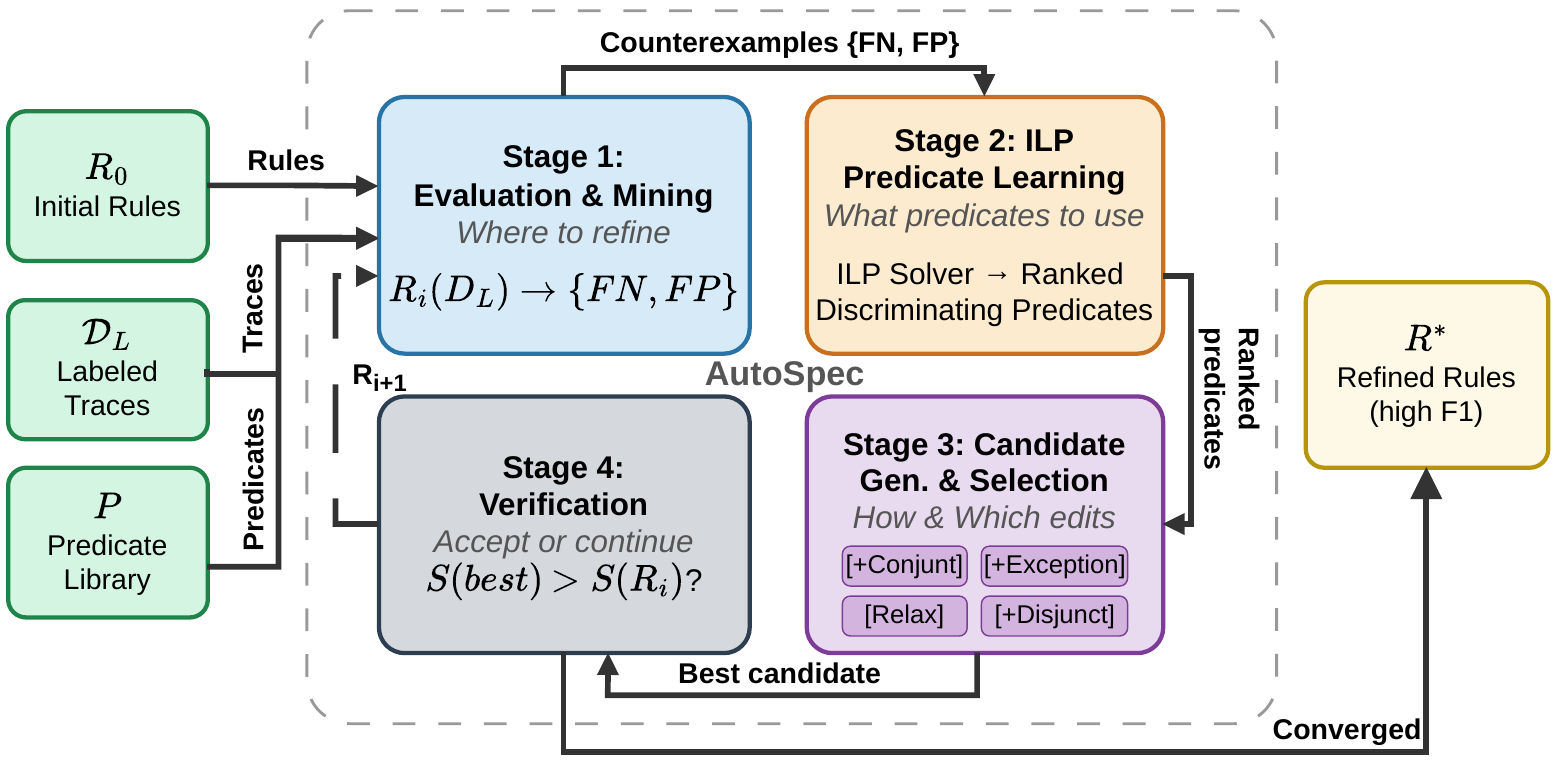}
    \caption{Overview of \tool.}
    \label{fig:workflow}
\end{figure}

\subsection{Workflow}
\label{sec:workflow}
{\F~\ref{fig:workflow}} illustrates \tool's architecture. The system takes as input: (1) an initial rule set $\mathcal{R}_0$, which in the intended deployment is the existing expert-authored guardrail for the domain (the algorithm also accepts a minimal seed rule when one wishes to bootstrap, but \tool's goal is to \emph{evolve} deployed rules, not to synthesize them from scratch), (2) a labeled dataset $\mathcal{D}_L$ of execution traces with ground-truth safety labels, and (3) a predicate library $\mathcal{P}$. \tool outputs a refined rule set $\mathcal{R}^*$ with high validation quality on $\mathcal{D}_L$, using F1 as the primary metric.

To efficiently navigate the massive edit space defined above, \tool organizes each refinement iteration into four focused stages. Each stage systematically narrows the search space for the next: counterexamples localize the failures, ILP selects the most discriminating predicates, constrained edit operators generate a manageable set of candidates, and a greedy-ascent check ensures monotonic progress. \A~\ref{alg:main} formalizes this loop, whose four stages we detail in \S~\ref{sec:approach} and summarize below:

\begin{algorithm}[t]
\caption{ILP-Guided Rule Evolution}
\label{alg:main}
\KwIn{Initial rules $\mathcal{R}_0$, labeled traces $\mathcal{D}_L$, predicate library $\mathcal{P}$, max iterations $K$, target score $\theta$}
\KwOut{Refined rules $\mathcal{R}^*$}
$\mathcal{R} \gets \mathcal{R}_0$\;
\For{$i = 1$ \KwTo $K$}{
    \lIf{$S(\mathcal{R}) \geq \theta$}{
        \Return $\mathcal{R}$;
    }
    $\mathit{CE} \gets \textsc{MineCounterexamples}(\mathcal{R}, \mathcal{D}_L)$\;
    \lIf{$\mathit{CE} = \emptyset$}{
        \Return $\mathcal{R}$;
    }
    $\mathit{suggestions} \gets \textsc{LearnILP}(\mathit{CE}, \mathcal{D}_L, \mathcal{P})$\;
    $\mathit{candidates} \gets \textsc{CandidateGen}(\mathcal{R}, \mathit{CE}, \mathit{suggestions})$\;
    $\mathit{best} \gets \textsc{SelectBest}(\mathit{candidates}, \mathcal{D}_L)$\;
    \lIf{$S(\mathit{best}) > S(\mathcal{R})$}{
        $\mathcal{R} \gets \mathit{best}$;
    }
    \lElse{
        \Return $\mathcal{R}$;
    }
}
\Return $\mathcal{R}$\;
\end{algorithm}

\ding{192}~\emph{Counterexample mining} evaluates $\mathcal{R}_i$ on $\mathcal{D}_L$ and collects the misclassified traces---false negatives (missed unsafe traces) and false positives (wrongly flagged safe traces)---that pinpoint where the rules fail. \ding{193}~\emph{ILP-guided predicate learning} finds the few predicates that best separate these counterexamples, shrinking the edit space to a short list of suggestions. \ding{194}~\emph{Candidate generation and selection} turns the suggestions into concrete edits with the operators of \T~\ref{tab:edit-ops}, scores each candidate on $\mathcal{D}_L$, and keeps the best. \ding{195}~\emph{Verification} accepts that candidate if it improves the score, and otherwise stops. The loop also ends when the target score is reached or the iteration budget $K$ is exhausted.

\subsection{Application Scope}
\label{sec:application-scope}

\begin{figure}[t]
    \centering
    \includegraphics[width=0.705\columnwidth]{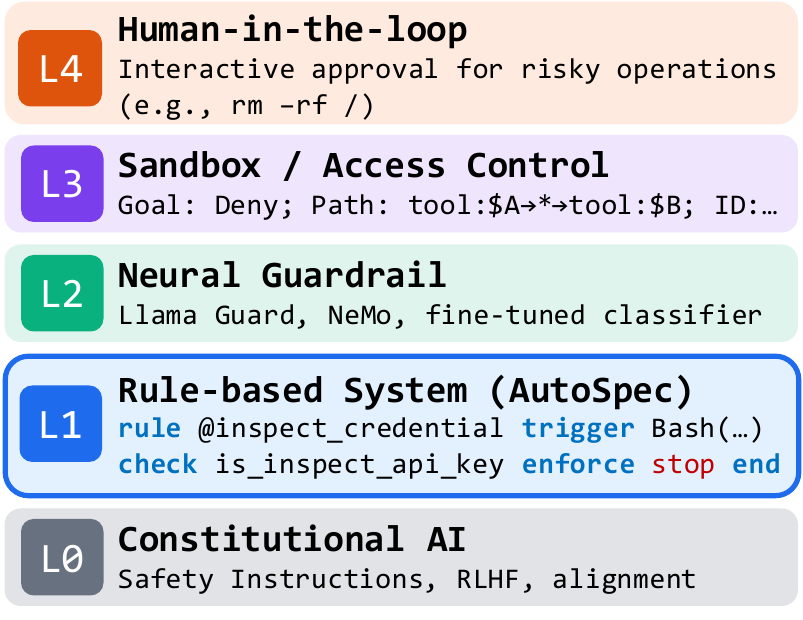}
    \caption{Defense-in-depth layers for LLM agent safety.}
    \label{fig:defense-layers}
\end{figure}

\tool targets the \emph{rule-based enforcement} layer in defense-in-depth architectures
for LLM agents (\F~\ref{fig:defense-layers}). Production deployments typically
combine multiple orthogonal layers: constitutional AI (Layer 0) for alignment, symbolic rules (Layer 1) for deterministic enforcement,
neural guardrails (Layer 2) for semantic detection, sandboxing (Layer 3) for execution isolation, and human supervision (Layer
4) for final verification.

While rule-based systems offer high explainability, their static nature makes them brittle against evolving threats. \tool addresses this by introducing ILP-guided CEGIS to automate rule refinement. This enables the symbolic layer to self-improve over time while preserving its core advantages: deterministic decisions, zero inference
cost, and auditability. Therefore, we anticipate \tool being most beneficial
when used in conjunction with other layers, where it can adapt the rule set to
evolving threats while leveraging the complementary strengths of the broader defense-in-depth framework.

We emphasize that \tool is intended for \emph{continuous, periodic maintenance} rather than one-shot offline tuning: deployed agents continually produce tool-call traces and approval signals that serve as labels, and \tool refines the deployed rules asynchronously as models, tools, and workloads evolve. In this setting ILP guidance contributes not merely search \emph{speed} but rule \emph{quality}: removing it lowers final quality substantially (heuristic CEGIS and random search reach far lower F1 than \tool in \S\ref{sec:evaluation}), because uniform exploration of the edit space rarely lands on the predicate combinations that cleanly separate counterexamples within a practical budget. We do not claim global optimality; the goal is practical, auditable improvement of the guardrails practitioners already run.

\section{Approach}
\label{sec:approach}
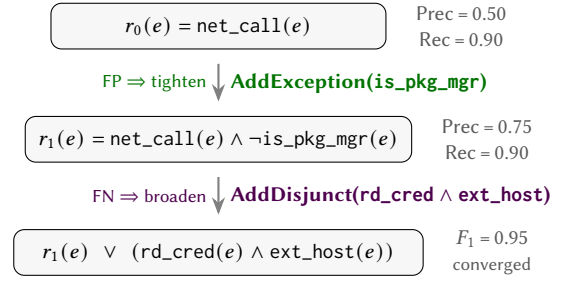
\begin{figure}[t]
\centering
\begin{tikzpicture}[
  font=\small,
  every node/.style={inner sep=0pt},
  rulebox/.style={draw, rounded corners=4pt, fill=gray!6, align=center, inner sep=5pt, font=\small, minimum width=14em},
  metbox/.style={font=\small\sffamily, text=gray!70!black, align=center},
  editlbl/.style={font=\small\sffamily\bfseries, text=violet!70!black, align=left, inner sep=0pt},
  arr/.style={thick, -{Stealth}, gray},
]

\node[rulebox] (r0) at (0, 0) {$r_0(e) = \texttt{net\_call}(e)$};
\node[metbox, right=1.2em of r0] {Prec\,=\,0.50\\ Rec\,=\,0.90};

\draw[arr] (0, -0.5) -- (0, -1.0);
\node[editlbl, right, text=green!45!black] at (0.15, -0.75) {AddException(\texttt{is\_pkg\_mgr})};
\node[anchor=east, font=\footnotesize\sffamily, text=green!45!black] at (-0.15, -0.75) {FP $\Rightarrow$ tighten};

\node[rulebox] (r1) at (0, -1.5) {$r_1(e) = \texttt{net\_call}(e) \land \neg\texttt{is\_pkg\_mgr}(e)$};
\node[metbox, right=1.2em of r1] {Prec\,=\,0.75\\ Rec\,=\,0.90};

\draw[arr] (0, -2.0) -- (0, -2.5);
\node[editlbl, right, text=violet!60!black] at (0.15, -2.25) {AddDisjunct(\texttt{rd\_cred} $\land$ \texttt{ext\_host})};
\node[anchor=east, font=\footnotesize\sffamily, text=violet!60!black] at (-0.15, -2.25) {FN $\Rightarrow$ broaden};

\node[rulebox, text width=16em] (r2) at (0, -3.0) {%
  $r_1(e) \;\lor\; (\texttt{rd\_cred}(e) \land \texttt{ext\_host}(e))$};
\node[metbox, right=1.2em of r2] {$F_1$\,=\,0.95\\ \footnotesize converged};

\end{tikzpicture}
\caption{Running example: a network-safety rule evolving over two CEGIS iterations---an FP-driven \textcolor{green!45!black}{tightening}, then an FN-driven \textcolor{violet!60!black}{broadening}.}
\label{fig:rule-evolution}
\end{figure}

We detail each stage of the loop below, following a single running example throughout (\F~\ref{fig:rule-evolution}). Starting from one rule $r_0(e) = \texttt{net\_call}(e)$ on a dataset of 10 unsafe and 15 safe traces, \tool reaches a precise rule in two iterations. \emph{Iteration~1} finds nine safe package-manager scripts wrongly flagged (FP), discovers that \texttt{is\_pkg\_mgr} distinguishes them, and adds it as an exception, raising precision to 0.75. \emph{Iteration~2} finds one missed exfiltration script (FN) that reaches the network through \texttt{subprocess}---so \texttt{net\_call} never fires---discovers that \texttt{rd\_cred} $\land$ \texttt{ext\_host} characterizes it, and adds that branch, at which point the rule set converges. The subsections that follow explain how each step is produced: counterexample mining (\S\ref{sec:counterexample-mining}) surfaces the FP and FN traces, the edit operators (\S\ref{sec:rule-edit-operations}) supply moves such as AddException and AddDisjunct, ILP (\S\ref{sec:ilp-guided-predicate-learning}) discovers the discriminating predicates, candidate generation and selection (\S\ref{sec:candidate-generation-verification}) pick the best edit, and the convergence rule (\S\ref{sec:multi-iteration-convergence}) decides when to stop.

\subsection{Rule Evaluation and Counterexample Mining}
\label{sec:counterexample-mining}

The first stage of each iteration evaluates the current rule set $\mathcal{R}_i$ on the labeled dataset $\mathcal{D}_L$ and extracts counterexamples that localize its failures.

\parh{Rule Evaluation.} For each trace $\tau \in \mathcal{D}_L$, \tool replays its events through $\mathcal{R}_i$ and labels $\tau$ \emph{unsafe} if any rule fires and \emph{safe} otherwise (\Def~\ref{def:safety-classification}). Comparing predictions with ground truth gives the usual true/false positives and negatives, and hence the precision, recall, and F1 of $\mathcal{R}_i$.

\parh{Counterexample Mining.} The misclassified traces are the engine of refinement, and the key observation is that each \emph{kind} of mistake points in a definite repair direction, as the two iterations of \F~\ref{fig:rule-evolution} show: a false negative (FN) is a missed unsafe trace, so the rule is too narrow and must be \emph{broadened}; a false positive (FP) is a false alarm, so the rule is too broad and must be \emph{tightened}. For every counterexample \tool records which rule fired and which predicate failed, handing the later stages not just a failing trace but the direction in which to repair it.

\subsection{Rule Edit Operations}
\label{sec:rule-edit-operations}

Given the counterexamples, \tool needs a vocabulary of rewrites that turn rules into more accurate variants. Each rule is a pair $r = (g, \phi)$: a trigger $g$ that scopes \emph{which} events the rule inspects (e.g., \texttt{before\_action} events on the \texttt{PythonREPL} tool), and a body $\phi$ that is a conjunction of predicates from $\mathcal{P}$ (\S~\ref{sec:background}). The operators in \T~\ref{tab:edit-ops} rewrite $\phi$ (single-rule edits) or the rule set $\mathcal{R}$ (disjunctive repair) while holding $g$ fixed, since in our domains the trigger is fixed by the guarded tool and is not a source of error; operators that also rewrite $g$ are a straightforward extension left to future work.

\begin{table}[h]
\centering
\caption{Primitive edit operators as rewrites over predicates.}
\label{tab:edit-ops}
\smallskip
\newcommand{\opspec}[1]{\colorbox{green!10}{\textcolor{green!45!black}{\small\sffamily\bfseries#1}}}
\newcommand{\opgen}[1]{\colorbox{violet!10}{\textcolor{violet!55!black}{\small\sffamily\bfseries#1}}}
\small
\begin{tabular}{@{}ll@{}}
\toprule
\textbf{Operator} & \textbf{Rewrite} \\
\midrule
\opspec{AddConjunct} & $(g,\;\phi) \;\Rightarrow\; (g,\;\phi \;\textcolor{green!45!black}{\land\; p})$ \\[0.3em]
\opspec{AddException} & $(g,\;\phi) \;\Rightarrow\; (g,\;\phi \;\textcolor{green!45!black}{\land\; \neg p})$ \\[0.3em]
\opgen{AddDisjunct} & $\mathcal{R} \;\Rightarrow\; \mathcal{R} \;\textcolor{violet!55!black}{\cup\; \{(g,\;\psi)\}}$ \\[0.3em]
\opgen{Relax} & $(g,\;\phi \;\textcolor{violet!55!black}{\land\; \ell}) \;\Rightarrow\; (g,\;\phi)$ \\
\bottomrule
\end{tabular}
\end{table}

\T~\ref{tab:edit-ops} summarizes the language. AddConjunct and AddException \emph{specialize} a rule (the latter via an explicit exception) to cut false positives; Relax does the inverse, dropping a literal to recover recall; AddDisjunct works at the rule-set level, appending a new branch to widen coverage without touching existing rules. Since a trace is unsafe whenever \emph{any} rule fires (\Def~\ref{def:safety-classification}), a two-branch set $\{r_1, (g,\psi)\}$ behaves as a disjunction, which we write as the shorthand $r_2(e) = r_1(e) \lor \psi(e)$. \tool searches the neighborhood these operators induce and validates candidates on labeled traces, as the example below illustrates.

\begin{example}
Consider a file-deletion safety rule that is iteratively refined.

\textbf{Baseline constraint.}
\[\small
r_0(e) = \texttt{destructive\_os\_inst}(e)
\]

\textbf{Exception edit.} If $r_0$ yields false positives on safe deletions of temporary files, \tool applies AddException and obtains
\[\small
r_1(e) = \texttt{destructive\_os\_inst}(e) \land \neg \textcolor{blue}{\texttt{is\_temp\_file}}(e)
\]

\textbf{Disjunctive repair.} If $r_1$ still misses unsafe deletions performed via \texttt{shutil.rmtree}, \tool adds a second branch:
\[\small
r_2(e) = r_1(e) \lor \textcolor{blue}{\texttt{uses\_shutil\_rmtree}}(e)
\]
This example shows how edit operators from \T~\ref{tab:edit-ops} can be composed to iteratively address different types of classification errors within a single refinement sequence.
\end{example}

\subsection{ILP-Guided Predicate Learning}
\label{sec:ilp-guided-predicate-learning}

The edit operators of \S\ref{sec:rule-edit-operations} still need to know \emph{which} predicates to apply. With 16--24 predicates per domain and several operators, trying every combination is infeasible, so \tool \emph{learns} the useful predicates from the counterexamples. The question it poses is deliberately narrow: among all predicates, which few separate the unsafe traces we missed (FN) from the safe traces we wrongly flagged (FP)? \tool answers it with Inductive Logic Programming (ILP)~\cite{muggleton1994inductive,law2020ilasp}, which (unlike a black-box classifier) returns a \emph{symbolic} hypothesis---a small combination of the existing predicates---and favors the fewest predicates, guarding against overfitting the handful of counterexamples. ILP does not write rules; it \emph{nominates predicates}, and the operators of \T~\ref{tab:edit-ops} do the rewriting.

\subsubsection{Learning Discriminating Predicates}

\tool poses this as an inductive learning task and hands it to an off-the-shelf ILP system, ILASP~\cite{law2020ilasp}.

\parh{Encoding.} To pose this question to the solver, \tool reuses the \emph{same} predicates, only evaluated per event: a fact $\texttt{holds}(\tau, i, p)$ records that predicate $p$ is true on event $i$ of trace $\tau$---no new representation is introduced. The missed unsafe traces (FN) are the positive examples and the wrongly flagged safe traces (FP) are the negatives, and \tool asks ILP for the \emph{simplest} predicate combination that holds on every positive but no negative. The answer has the form
\[
H = (p_{11} \land p_{12} \land \cdots) \;\lor\; \cdots \;\lor\; (p_{k1} \land p_{k2} \land \cdots),
\]
an \emph{or} of small \emph{and}-groups, where the predicates in a group must hold together on one event. ``Simplest'' means fewest predicates, which keeps the result readable. The predicates appearing in $H$ are exactly the discriminating ones \tool turns into rule edits, as the example illustrates.

\begin{example}
Zooming into iteration~2 of the running example (\F~\ref{fig:rule-evolution}), the current rule flags any trace containing a network call, producing two FP traces $\tau_{\text{fp1}}, \tau_{\text{fp2}}$ (safe scripts fetching package metadata) and one FN trace $\tau_{\text{fn}}$ (credential exfiltration). The predicate library includes \texttt{net\_call}, \texttt{reads\_cred}, \texttt{ext\_host}, and \texttt{is\_pkg\_mgr}. The diagram below shows which predicates hold on each trace (top) and the clause the ILP solver learns (bottom).

\smallskip
\begin{center}
\resizebox{\columnwidth}{!}{%
\begin{tikzpicture}[
  font=\small\ttfamily,
  every node/.style={inner sep=0pt},
  hdr/.style={font=\small\sffamily\bfseries, text height=1.4ex, text depth=0.3ex},
  cell/.style={minimum width=1.2em, minimum height=1.15em, anchor=center},
  yes/.style={cell, text=green!50!black, font=\small\bfseries},
  no/.style={cell, text=red!70!black, font=\small},
  lbl/.style={anchor=east, font=\small\ttfamily},
  tag/.style={rounded corners=2pt, font=\footnotesize\sffamily, inner sep=1.5pt, minimum height=1.2em},
  tagfn/.style={tag, fill=red!12, text=red!60!black},
  tagfp/.style={tag, fill=blue!10, text=blue!50!black},
]

\def\colA{0}
\def\colB{1.75}
\def\colC{3.5}
\def\colD{5.25}

\draw[gray!40] (-0.875, -0.2) -- (6.125, -0.2);
\draw[gray!40] (-0.875, -0.6) -- (6.125, -0.6);
\draw[gray!40] (-0.875, -1.0) -- (6.125, -1.0);
\draw[gray!40] (-0.875, -1.4) -- (6.125, -1.4);
\draw[gray!40] (-0.875, 0.26) -- (-0.875, -1.4);
\draw[gray!40] (0.875, 0.26) -- (0.875, -1.4);
\draw[gray!40] (2.625, 0.26) -- (2.625, -1.4);
\draw[gray!40] (4.375, 0.26) -- (4.375, -1.4);
\draw[gray!40] (6.125, 0.26) -- (6.125, -1.4);
\draw[gray!60] (-0.875, 0.26) -- (6.125, 0.26);

\node[hdr] at (\colA, 0) {net\_call};
\node[hdr] at (\colB, 0) {rd\_cred};
\node[hdr] at (\colC, 0) {ext\_host};
\node[hdr] at (\colD, 0) {pkg\_mgr};

\node[lbl] at (-1.0, -0.4) {$\tau_{\text{fn}}$};
\node[tagfn] at (-1.85, -0.4) {FN};
\node[no, cell] at (\colA, -0.4) {--};
\node[yes] at (\colB, -0.4) {\cmark};
\node[yes] at (\colC, -0.4) {\cmark};
\node[no, cell] at (\colD, -0.4) {--};

\node[lbl] at (-1.0, -0.8) {$\tau_{\text{fp1}}$};
\node[tagfp] at (-1.85, -0.8) {FP};
\node[yes] at (\colA, -0.8) {\cmark};
\node[no, cell] at (\colB, -0.8) {--};
\node[no, cell] at (\colC, -0.8) {--};
\node[yes] at (\colD, -0.8) {\cmark};

\node[lbl] at (-1.0, -1.2) {$\tau_{\text{fp2}}$};
\node[tagfp] at (-1.85, -1.2) {FP};
\node[yes] at (\colA, -1.2) {\cmark};
\node[no, cell] at (\colB, -1.2) {--};
\node[no, cell] at (\colC, -1.2) {--};
\node[yes] at (\colD, -1.2) {\cmark};

\draw[thick, -{Stealth}, gray] (2.625, -1.58) -- node[right=2pt, font=\footnotesize\sffamily, text=gray]{ILP} (2.625, -1.94);

\node[draw, rounded corners=3pt, fill=gray!6,
      align=center, anchor=north,
      font=\small, inner sep=4pt]
  at (2.625, -2.02) {%
    learns\ \ $\texttt{rd\_cred} \land \texttt{ext\_host}$
  };

\end{tikzpicture}%
}
\end{center}
\smallskip

\noindent The learned clause covers $\tau_{\text{fn}}$ (both predicates hold, at different steps) without covering the FP traces, identifying \texttt{reads\_cred} and \texttt{ext\_host} as the discriminating predicates for candidate generation.
\end{example}

\subsubsection{From Hypothesis to Edit Suggestions}

Each predicate in $H$ carries a direction, which selects its operator just as in the two iterations of \F~\ref{fig:rule-evolution}: a predicate that helps \emph{cover} a missed unsafe trace raises recall (applied via AddDisjunct or Relax), whereas one that helps \emph{exclude} a false alarm raises precision (AddConjunct or AddException). \tool scores each suggestion by how cleanly it separates the counterexamples---the FN traces it newly covers minus the FP traces it newly violates---and then applies a greedy set cover, with separate quotas for recall- and precision-improving operators, to keep a small, diverse batch instead of many variants of the same fix. Every surviving suggestion becomes a candidate that is verified on $\mathcal{D}_L$ (\S\ref{sec:candidate-generation-verification}); the scores only steer exploration and never replace this verification.

\subsection{Candidate Generation and Selection}
\label{sec:candidate-generation-verification}

From the ILP-pruned suggestions, \tool builds candidate rule sets through three strategies: a \emph{single-step} edit applies one suggestion to the implicated rule; a \emph{mixed repair chain} bundles a precision-improving edit with one or more recall-improving ones, for when fixing an FN creates new FPs; and a \emph{failure-driven branch} adds high-precision disjuncts for residual FNs via a precision-constrained greedy set cover. \tool then evaluates each candidate on $\mathcal{D}_L$ and greedily keeps, as $\mathcal{R}_{i+1}$, the one maximizing $S(\mathcal{R}') = (F_1,\, \mathrm{Prec},\, -U,\, -|\mathcal{R}'|)$ in lexicographic order (F1, then precision, then fewer unresolved counterexamples $U$, then fewer rules).

\subsection{Multi-Iteration Convergence}
\label{sec:multi-iteration-convergence}

The greedy ascent in \A~\ref{alg:main} improves the score each iteration, but across iterations the ILP solver tends to re-propose predicates already in the rule set (they still look discriminative), which stalls progress. \tool down-weights suggestions that overlap the current rules, steering each iteration toward novel refinements while still allowing reuse when the score gain is large enough---a soft novelty pressure that prevents oscillation between near-identical rule sets.

The loop terminates when the current rule set reaches the target score threshold, when no candidate improves upon the current score, or when the iteration budget is exhausted. Iteration~2 of the running example (\F~\ref{fig:rule-evolution}) exercises this diversity pressure: \texttt{rd\_cred} and \texttt{ext\_host} are down-weighted because they appeared in the previous hypothesis, yet they are still selected since no alternative matches their coverage.

\parh{Convergence properties.} The algorithm terminates in finite time (bounded
by $K$ iterations) with three guarantees: (1) monotonic score improvement
($S(\mathcal{R}_{i+1}) \geq S(\mathcal{R}_i)$), (2) local optimality at
termination (no single-step edit improves the score), and (3) deterministic
output given fixed inputs. However, since the underlying synthesis problem is
NP-hard, we do not guarantee global optimality or convergence rate; solution
quality depends on the richness of the predicate library $\mathcal{P}$. Despite
that, we find that even a modest predicate set combined with ILP guidance can
yield significant improvements in practice, as shown in our experiments.

%% file: setup.tex
\section{Experimental Setup}
\label{sec:setup}

\subsection{Research Questions}

We evaluate \tool to answer the following research questions (RQs):

\begin{description}[leftmargin=0pt]
\item[RQ1: Effectiveness] How much does \tool improve safety rule quality (precision, recall, F1) compared to baseline rules?

\item[RQ2: Search Efficiency] How efficiently does \tool refine rules, in terms of convergence behavior and the contribution of ILP guidance?

\item[RQ3: Transfer and Usability] Do the learned rules generalize across agent configurations, and do practitioners find them interpretable and actionable?
\end{description}

\subsection{Datasets}

We evaluate \tool on execution traces from the two application domains introduced in \S\ref{sec:background}: code execution and embodied household robotics. All traces are manually labeled as safe or unsafe at trace granularity as defined in \S\ref{sec:background}. Representative predicates for these domains are shown in \T~\ref{tab:predicates}, and the full domain-specific predicate libraries are deferred to the supplementary material.

\textbf{Code Execution Domain.} Following the tool-using LLM-agent setting discussed in \S\ref{sec:background} and implemented in AgentSpec~\cite{wang2025agentspec}, we collect traces from agents invoking PythonREPL. Unsafe traces (91) are sampled from RedCode-Exec~\cite{guo2024redcode}, covering 25 risk categories (destructive operations, sensitive file access, dangerous subprocess calls, unauthorized network requests), with up to 4 traces per category. Safe traces (100) are generated using GPT-5.1-Codex with prompts producing benign operations that mirror unsafe categories but avoid triggering security predicates. All 191 traces are labeled and used for rule synthesis and evaluation. Code-execution traces are short, ranging from 1 to 9 events (mean 1.4).

\textbf{Embodied Agent Domain.} Following the embodied household-robot setting in \S\ref{sec:background} and prior work on language-conditioned robot planning~\cite{ahn2022can}, we collect traces from SafeAgentBench~\cite{yin2024safeagentbench}, where simulated household robots in AI2-THOR perform tasks like ``fetch the laptop from the table.'' Agents use GPT-5.1-Codex with zero-shot ReAct prompting and interact through a high-level robotic controller supporting 16 primitive actions (e.g., find, pick, slice). Traces are labeled \emph{unsafe} when the robot approaches danger zones without precautions, handles fragile objects with excessive force, carries liquids near electronics, or ignores collision risks. Safe traces involve legitimate household tasks with proper object handling. This dataset contains 100 traces (52\% unsafe); embodied traces are longer than code-execution traces, ranging from 1 to 15 events (mean 5.1).

\subsection{Evaluation Metrics}

Because rules fire on individual events but labels are provided per trace (see \S\ref{sec:background}), all classification metrics are reported at trace granularity: a trace is predicted unsafe if any rule fires on any of its events, and safe otherwise.

We report standard classification metrics: precision, recall, and F1 score.
We also measure convergence iterations (the number of CEGIS iterations until termination), rule complexity (the number of rules and predicates in the final rule set), and candidate efficiency (the fraction of generated candidates that improve F1).

\subsection{Baselines}

We compare \tool against the following baselines. As discussed in \S~\ref{sec:application-scope}, we focus on approaches within the rule-based layer rather than cross-layer alternatives (e.g., neural guardrails, sandboxing), as the latter would conflate orthogonal design axes.

\textbf{B1: Expert-Designed Rule.} The expert-authored
AgentSpec~\cite{wang2025agentspec} guardrail deployed for the domain, i.e., the
human-written, expert-optimized rule set rather than a toy rule authored only for
this evaluation. B1 therefore \emph{is} our comparison against expert-designed
rules: it is simultaneously the starting point for evolution and the human-expert
reference. This reflects \tool's positioning, which is not to \emph{reinvent}
safety rules from scratch but to \emph{evolve} expert-written static rules as the
agent environment, models, tools, and human annotations shift. Accordingly, the
margin of \tool over B1 in \T~\ref{tab:effectiveness} directly quantifies the
value \tool adds on top of expert-designed rules.

\textbf{B2: LLM-based Classifier.} A neural classifier that consumes the
\emph{same} labeled traces as \tool, but uses them as few-shot exemplars (5 per
class) to label a trace safe or unsafe rather than to evolve rules. It is the
alternative way of turning the annotation feedback into an adaptive guardrail.
Our claim is not universal superiority over every LLM classifier, but that
\tool turns the same signal into deterministic, auditable rules with strong
effectiveness and better actionability.

\textbf{B3: CEGIS without ILP.} \tool without ILP guidance, where candidate generation uses only coverage-based predicate ranking without learning from counterexamples.

\textbf{B4: Random Search.} Randomly sample candidate edits from the search space and select the best after a fixed budget of evaluations.

\subsection{Experimental Protocol}
\label{sec:protocol}

We make the measurement protocol explicit because two of our research questions
report scores under different conditions. In \textbf{RQ1} (\T~\ref{tab:effectiveness}),
the refinement methods (B3, B4, and \tool) evolve the initial guardrail on a
\emph{seed} split and are then evaluated on a disjoint \emph{held-out} test
split, while B1 is that same guardrail evaluated \emph{without} refinement. The
greedy-acceptance step in \A~\ref{alg:main} guarantees monotonic improvement on
the \emph{seed} score, not on the held-out score; consequently a refined method
can score below B1 on the held-out split when it over-specializes to the seed.
This is a generalization gap, not a violation of Algorithm~\ref{alg:main}, and is
precisely where ILP helps (\S\ref{sec:evaluation}). In \textbf{RQ2}
(\F~\ref{fig:convergence}, \T~\ref{tab:ilp-impact}), we instead report the
\emph{in-sample} convergence trajectory on the refinement set; for the code
domain the curve is seeded from a minimal single-predicate rule (hence
F1~$\approx$~0 at iteration~0), whereas the embodied domain is seeded from the
initial guardrail (F1~$=$~0.493). \textbf{RQ3} uses a 70\%/30\% split into
initial and incoming tasks: rules are learned on the initial split and evaluated
on the disjoint held-out incoming split, which simulates the future workloads the
rules must generalize to.

\subsection{Implementation}

\tool is implemented in Python 3.10 with 12K lines of code (on the basis of
AgentSpec's predicate library). We use ILASP v4.4.1~\cite{law2020ilasp} as the
ILP backend with a 120-second timeout per learning task. All experiments run on
a MacBook Pro (M2 Max, 32 GB RAM).

%% file: evaluation.tex
\section{Evaluation}
\label{sec:evaluation}

\subsection{RQ1: Effectiveness}

\begin{table}[ht]
\centering
\caption{Effectiveness comparison across domains. Best results in \textbf{bold}.}
\label{tab:effectiveness}
\small
\resizebox{0.9\columnwidth}{!}{
\begin{tabular}{@{}lcccc@{}}
\toprule
\textbf{Method} & \textbf{Precision} & \textbf{Recall} & \textbf{F1} & \textbf{FP/FN} \\
\midrule
\multicolumn{5}{l}{\textit{Code Execution Domain}} \\
Baseline Rule (B1) & 0.544 & 0.974 & 0.698 & 31/1 \\
LLM Classifier (B2) & 1.000 & 0.684 & 0.813 & 0/12 \\
CEGIS w/o ILP (B3) & 1.000 & 0.316 & 0.480 & 0/26 \\
Random Search (B4) & 0.536 & 0.974 & 0.692 & 32/1 \\
\tool (Ours) & \textbf{0.960} & \textbf{1.000} & \textbf{0.980} & \textbf{2/0} \\
\midrule
\multicolumn{5}{l}{\textit{Embodied Agent Domain}} \\
Baseline Rule (B1) & 1.000 & 0.327 & 0.493 & 0/35 \\
LLM Classifier (B2) & 0.842 & 0.923 & 0.881 & 9/4 \\
CEGIS w/o ILP (B3) & 1.000 & 0.500 & 0.667 & 0/26 \\
Random Search (B4) & 1.000 & 0.327 & 0.493 & 0/35 \\
\tool (Ours) & \textbf{0.925} & \textbf{0.942} & \textbf{0.933} & \textbf{4/3} \\
\bottomrule
\end{tabular}
}
\end{table}

\T~\ref{tab:effectiveness} presents the effectiveness of \tool compared to baselines across two domains. The results demonstrate that \tool consistently achieves the highest F1 scores across both domains, with F1 of 0.980 for code execution and 0.933 for embodied agents.

Among the baselines, the expert-designed rules (B1) exhibit contrasting failure patterns across domains: over-approximation in code execution (recall 0.974 but precision 0.544, producing 31 FP) versus under-approximation in embodied agents (precision 1.000 but recall 0.327, missing 35 unsafe cases). This illustrates the fundamental precision-recall tradeoff of static rules. The LLM classifier (B2) achieves moderate F1 (0.813 and 0.881) but struggles to capture precise decision boundaries, missing 12 unsafe cases in code execution and producing 9 false positives in embodied agents; moreover, its black-box nature limits interpretability and actionability. CEGIS without ILP (B3) maintains perfect precision but suffers from extremely low recall (0.316 and 0.500), missing 26 unsafe cases in each domain, demonstrating that heuristic predicate selection fails to discover discriminating predicates. Notably, in code execution B3's held-out F1 (0.480) falls \emph{below} the unrefined baseline B1 (0.698). Per the protocol in \S\ref{sec:protocol}, this is a generalization gap rather than a contradiction of Algorithm~\ref{alg:main}: without ILP guidance B3 over-specializes to the seed split (seed precision 1.0, recall 0.316), so its held-out F1 drops; \tool, refined from the same seed, instead reaches 0.980, directly isolating ILP's contribution. Random search (B4) performs comparably to B1, confirming that guided search is essential.
In contrast, \tool achieves near-optimal performance: F1 of 0.980 in code execution (only 2 FP, perfect recall) and 0.933 in embodied agents (4 FP, 3 FN), improving over baselines by 40.4\% and 89.2\% respectively. These results demonstrate that ILP-guided CEGIS effectively learns precise safety rules that balance false positive and false negative rates.

\subsection{RQ2: Search Efficiency}

\subsubsection{Convergence}

\begin{figure}[t]
    \centering
    \includegraphics[width=0.99\linewidth]{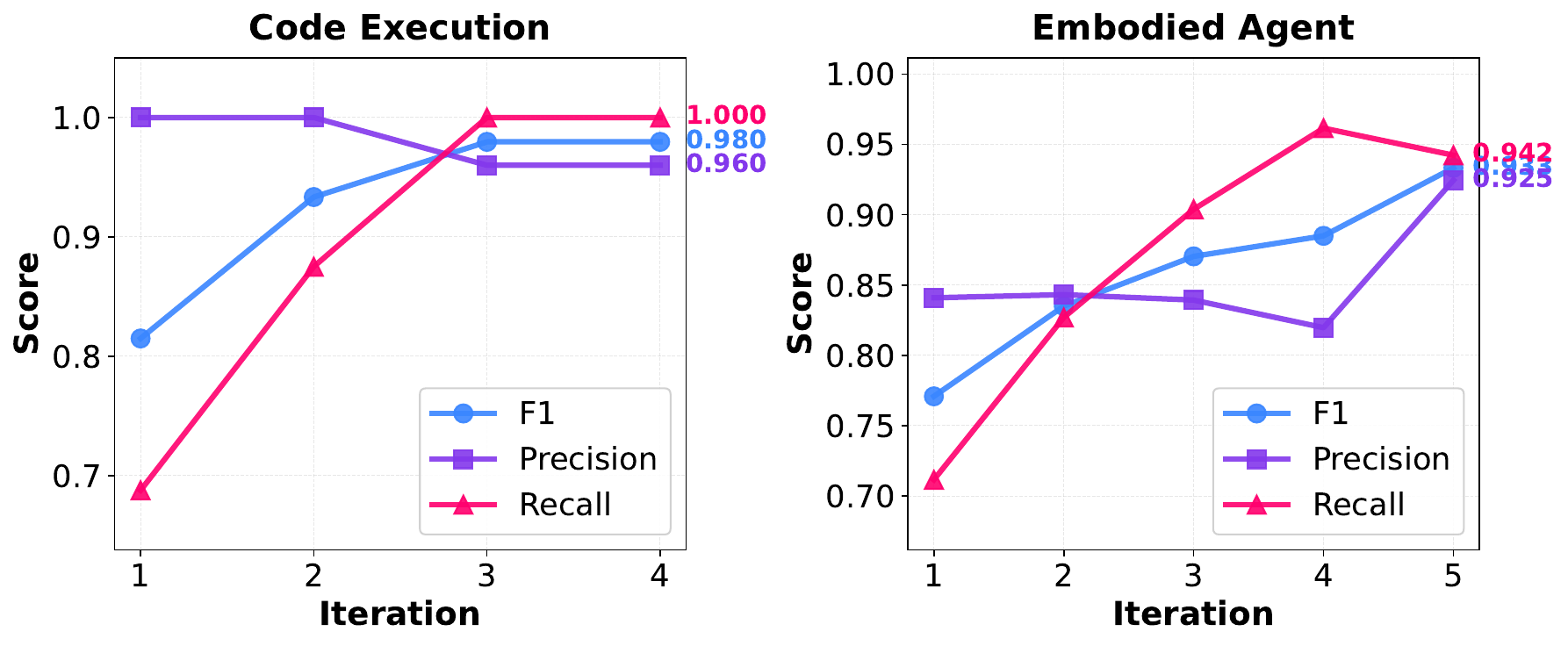}
    \caption{Convergence trajectories showing F1, Precision, and Recall over iterations. Code execution converges in 4 iterations (F1=0.980), while embodied agents require 5 iterations (F1=0.933).}
    \label{fig:convergence}
\end{figure}

\F~\ref{fig:convergence} shows that \tool achieves rapid convergence in both domains. In code execution, \tool reaches the target F1 threshold of 0.95 within 4 iterations (final F1 0.980), with the first iteration alone jumping from 0.000 to 0.815, indicating that the predicate library contains highly discriminating features. The 0.000 starting point reflects the RQ2 protocol (\S\ref{sec:protocol}): the code-domain convergence curve is seeded from a minimal single-predicate rule that fires on nothing, which is distinct from the deployed guardrail used as B1 in RQ1 (F1 0.698). This is a deliberate stress test of convergence from a weak seed, not \tool's intended deployment mode; in deployment \tool evolves the expert guardrail, as in the embodied curve below, which starts from that guardrail (F1 0.493), and as in RQ1. The embodied agent domain converges more gradually: starting from F1 of 0.493, the system improves through 5 iterations (0.493 → 0.771 → 0.835 → 0.870 → 0.885 → 0.933), terminating at the iteration limit rather than the target threshold. The final iteration still shows meaningful improvement (0.885 to 0.933), suggesting additional iterations could yield further gains.

\subsubsection{Impact of ILP Guidance}

\T~\ref{tab:ilp-impact} compares \tool with ILP-guided predicate learning against a variant using heuristic-based predicate ranking. Both variants learn from counterexamples, but differ in how they prioritize predicates for candidate generation. We emphasize that \tool~(Heuristic) here and CEGIS~w/o~ILP~(B3) in \T~\ref{tab:effectiveness} are the \emph{same} ablation, i.e., \tool with ILP removed so that candidate predicates are ranked by coverage statistics instead of ILP suggestions (selection is deterministic, not random sampling). They report different F1 only because of the protocol in \S\ref{sec:protocol}: B3 is measured on the held-out split, whereas \tool~(Heuristic) reports the in-sample RQ2 convergence trajectory.

\begin{table}[t]
\centering
\caption{ILP-guided vs heuristic predicate selection.}
\label{tab:ilp-impact}
\small
\resizebox{\columnwidth}{!}{
\begin{tabular}{@{}lcccc@{}}
\toprule
\textbf{Method} & \textbf{Iterations} & \textbf{Final F1} & \textbf{Candidates} & \textbf{Time (s)} \\
\midrule
\multicolumn{5}{l}{\textit{Code Execution Domain}} \\
\tool (Heuristic) & 2 & 0.203 & 197 & 4.4 \\
\tool (ILP) & 4 & 0.980 & 319 & 20.1 \\
\midrule
\multicolumn{5}{l}{\textit{Embodied Agent Domain}} \\
\tool (Heuristic) & 4 & 0.758 & 372 & 4.3 \\
\tool (ILP) & 5 & 0.933 & 600 & 11.8 \\
\bottomrule
\end{tabular}
}
\end{table}

The results demonstrate that ILP-guided learning significantly outperforms heuristic approaches in both domains. In code execution, it achieves F1 of 0.980 versus the heuristic's 0.203 (a 382\% improvement); the heuristic terminates after only 2 iterations, unable to find improving candidates without formal learning from counterexamples. In the embodied agent domain, ILP achieves F1 of 0.933 versus 0.758 (23\% improvement). The heuristic evaluates 372 candidates versus ILP-guided learning's 600, but the additional candidates explored by ILP are better targeted due to formal learning, leading to superior final quality.

Regarding the execution time, ILP-guided learning incurs moderate time overhead (20.1s vs 4.4s for code, 11.8s vs 4.3s for embodied), but the heuristic approach produces rules inadequate for deployment. Given the substantial quality gains, this overhead is a favorable tradeoff, especially as \tool can operate asynchronously alongside agent execution without increasing actual user wait time in practice.

\subsection{RQ3: Transfer and Usability}

\subsubsection{Generalization}

\begin{figure}[t]
    \centering
    \includegraphics[width=0.99\linewidth]{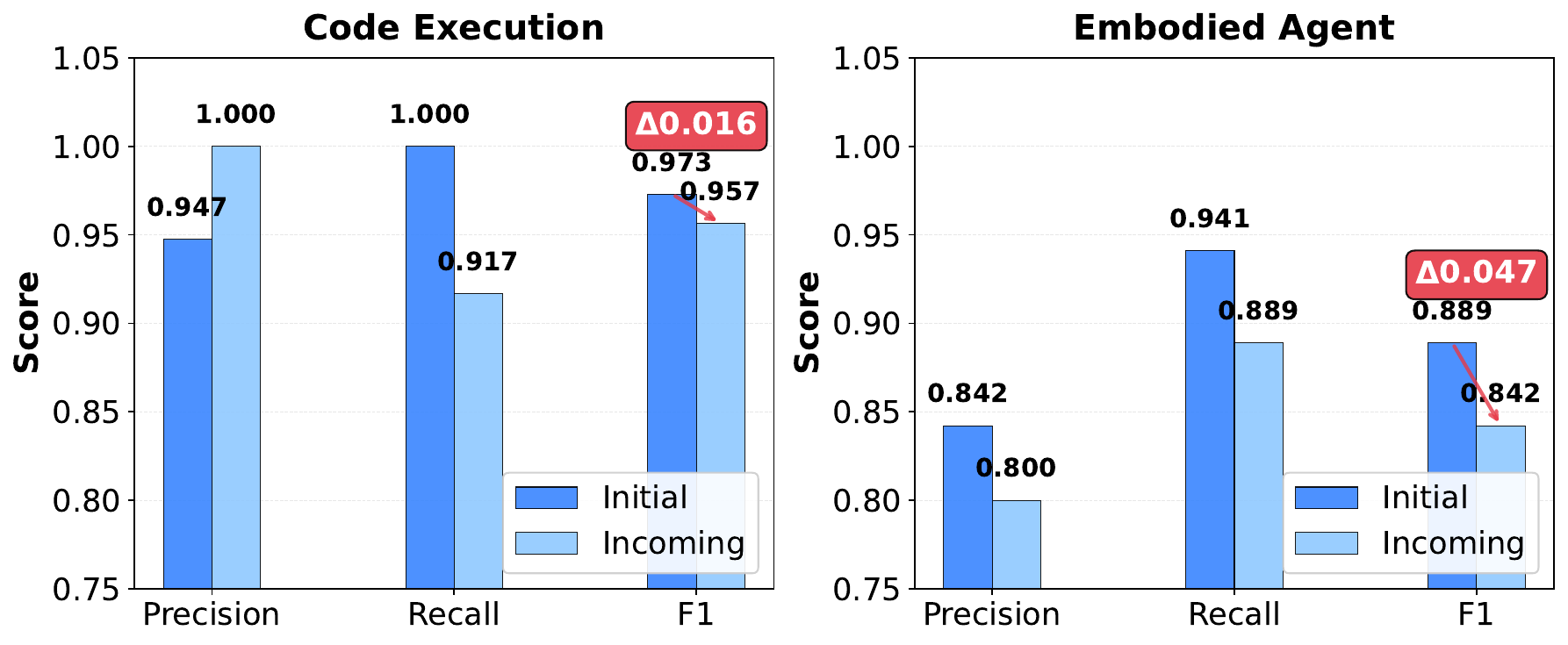}
    \caption{Generalization to incoming agent tasks. Rules learned from 70\% of initial tasks maintain high performance on 30\% incoming tasks, with small generalization gaps of 1.6\% (code) and 4.7\% (embodied).}
    \label{fig:generalization}
\end{figure}

We evaluate whether rules learned from initial agent tasks generalize to incoming tasks. For each domain, we randomly partition the dataset into 70\% initial tasks (used for rule learning) and 30\% incoming tasks (held out to simulate future workloads). \F~\ref{fig:generalization} visualizes the performance comparison.

\F~\ref{fig:generalization} shows that \tool demonstrates strong generalization across both domains, with minimal performance degradation on incoming tasks. In code execution, the learned rules achieve F1 of 0.973 on initial tasks and 0.957 on incoming tasks (1.6 percentage point gap), with incoming tasks achieving perfect precision (1.000) and high recall (0.917), indicating the rules capture genuine unsafe patterns rather than memorizing examples. In the embodied agent domain, the gap is slightly larger (F1 0.889 → 0.842, 4.7 points), with balanced precision (0.800) and recall (0.889) on incoming tasks. Overall, these results confirm that ILP-guided CEGIS produces rules that generalize well to evolving agent workloads, a critical property for production deployment.

\subsubsection{Interpretability}

To evaluate whether \tool's learned rules are interpretable to practitioners, we conducted a user study comparing \tool's learned rules against LLM classifier decisions.

\parh{Choice of baseline.} Accumulated annotations can be turned into an adaptive
guardrail in two ways: by training a neural classifier over the labeled traces,
or by evolving symbolic rules from them, as \tool does. Both consume the same
feedback and differ only in the form of their output, opaque scores versus
inspectable rules, so they are the two options a practitioner actually chooses
between. We therefore compare them on the properties that decide deployment:
interpretability, actionability, and trust. The static expert guardrail is
\tool's input rather than an adaptive alternative, and shares \tool's symbolic
form, so it is not the relevant contrast here.

\textbf{Protocol.} We recruited 10 practitioners with more than one year of
experience in developing or deploying ML-based systems. Participants evaluated
two approaches across four scenarios (two code execution, two embodied agent
domains). For each scenario, participants were shown: (1) an execution trace,
(2) the ground truth label, and (3) the decision explanation from each method.
Participants rated each method on three dimensions using a 5-point Likert scale
(1=Very Poor, 5=Excellent): \textbf{Interpretability} (ease of understanding why
a decision was made), \textbf{Actionability} (ease of modifying the approach to
improve it), and \textbf{Trust} (confidence in deploying the approach in
production). Participants were also encouraged to provide brief written explanations for their ratings and suggested modifications to each approach.

\begin{figure}[t]
    \centering
    \includegraphics[width=0.75\linewidth]{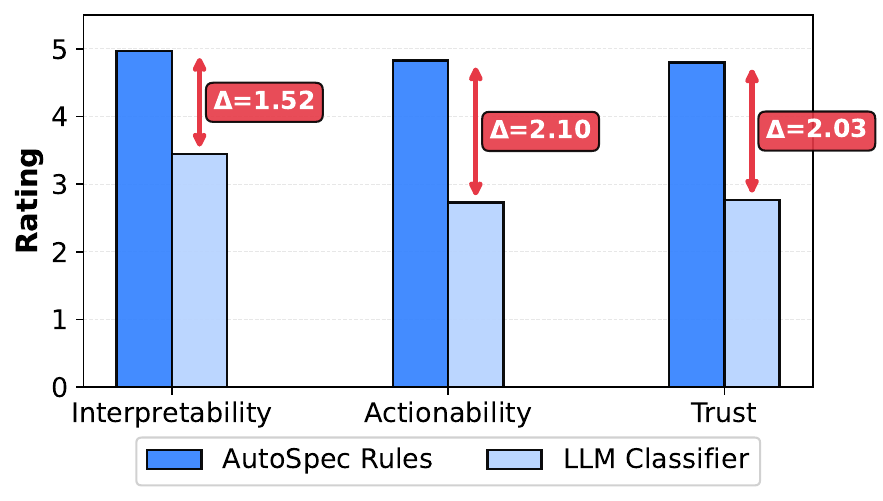}
    \caption{Interpretability evaluation comparing \tool rules and LLM classifiers across three dimensions.}
    \label{fig:interpretability}
\end{figure}

\textbf{Results.} As \F~\ref{fig:interpretability} summarizes, \tool rules
outperformed LLM classifiers on all three dimensions (Wilcoxon signed-rank test,
$p<0.01$), and participants preferred \tool in 101 of 120 pairwise comparisons
(84.2\%), the remaining 19 being ties. The per-dimension gaps reveal \emph{where}
symbolic rules help most. On \textbf{interpretability} (4.97 vs.\ 3.45),
participants found \tool rules fully traceable, pinpointing which predicate fired
and reasoning about counterfactuals, whereas LLM explanations read fluently yet
left the decision boundary opaque. On \textbf{actionability} (4.83 vs.\ 2.73, the
widest gap), they proposed exact, locally scoped edits to \tool rules but doubted
prompt-level fixes to the classifier: ``we can not ensure LLM not to accidentally
unblock those calls by these few changes.'' On \textbf{trust} (4.80 vs.\ 2.77),
all 10 ranked \tool most production-ready, citing determinism and auditability,
while the classifier was seen as ``prone to hallucinations.'' Notably, the
actionability and trust gaps (both $>2$ points) far exceed the interpretability
gap ($1.5$): practitioners could \emph{read} the LLM's rationale but not safely
\emph{act on} or \emph{rely on} it, which is exactly where \tool's symbolic
output pays off.

\parh{Takeaway.} The rules participants judged were \tool's synthesized
constructs, the predicate combinations, exceptions, and disjunctive branches it
evolved from counterexamples, so the ratings reflect \tool's actual output. The
picture is consistent: a neural classifier learns from the same annotations but
produces decisions practitioners find hard to inspect, edit, and trust, whereas
\tool reaches comparable adaptivity while staying auditable. Adapting from
feedback without giving up interpretability is the combination \tool is designed
to provide.

\subsection{Ablation Study}

\T~\ref{tab:ablation} presents an ablation study evaluating the contribution of key components in \tool. The results validate the design choices made in \S~\ref{sec:ilp-guided-predicate-learning} and \S~\ref{sec:candidate-generation-verification}, demonstrating that ILP guidance is the most critical component, with its removal causing F1 to drop from 0.98 to 0.20. Multi-step edits and failure-driven branches also contribute significantly to both effectiveness and efficiency.

\begin{table}[h]
\centering
\caption{Ablation study on code execution domain.}
\label{tab:ablation}

\begin{tabular}{@{}lccc@{}}
\toprule
\textbf{Variant} & \textbf{F1} & \textbf{Iterations} & \textbf{Candidates} \\
\midrule
\tool (Full) & 0.98 & 4 & 319 \\
- ILP guidance & 0.20 & 2 & 197 \\
- Failure-driven branches & 0.97 & 6 & 677 \\
- Multi-step edits & 0.72 & 10 & 300 \\
\bottomrule
\end{tabular}
\end{table}

\textbf{Finding 1: ILP guidance is essential.} Removing ILP-based predicate learning causes F1 to drop by 78 points (from 0.98 to 0.20), confirming that the ILP approach in \S~\ref{sec:ilp-guided-predicate-learning} is critical for identifying discriminating predicates. Without ILP, the system falls back to heuristic predicate ranking, which fails to learn which predicates characterize unsafe traces.

\textbf{Finding 2: Multi-step edits handle complex interactions.} Restricting the system to single-step edits reduces F1 to 0.72 and increases iterations from 4 to 10, demonstrating why the mixed repair chains in \S~\ref{sec:candidate-generation-verification} are necessary: many counterexamples require coordinated changes across multiple rule components. This is a factor distinct from ILP: repair chains form whether predicates are ranked by ILP or by coverage, so the two rows ablate independent components.

\textbf{Finding 3: Failure-driven branches improve efficiency.} Disabling failure-driven disjunct generation maintains high F1 (0.97) but doubles the iteration count (from 4 to 6) and increases candidate evaluations by 112\% (from 319 to 677). The precision-constrained greedy set cover strategy (\S~\ref{sec:candidate-generation-verification}) accelerates convergence by directly generating high-precision disjunctive branches without introducing excessive false positives.

%% file: discussion.tex
\section{Discussion and Threats to Validity}
\label{sec:discussion}

\parh{Handling agent-specific safety concerns.}
Open-ended tool use, hallucinated actions, context-dependent safety, and
adversarial manipulation are all handled at the \emph{predicate layer}:
predicates expose tool calls, action history, state snapshots, and domain
context, and integrating a new tool simply adds predicates through the existing
AgentSpec interface. \tool evolves executable, auditable guardrails over these
predicates, and adversarial traces obtained from red-teaming~\cite{perez2022red}
enter the loop as additional counterexamples, so coverage strengthens as new
attack patterns are observed. Because predicates are the unit of expressiveness,
the counterexamples that remain unresolved at convergence also pinpoint exactly
where a new predicate would help, turning library coverage into a measurable
signal rather than a hidden assumption.

\parh{Design choices.}
\tool drives candidate generation with ILP rather than a decision tree or a
single global synthesis call by design. A decision tree classifies feature
vectors, whereas \tool must emit AgentSpec rules with triggers, enforcement
actions, and localized, traceable edits to a deployed rule set, for which ILP
provides a simplicity bias under mode declarations that a greedy split lacks. The
CEGIS loop further passes only the current counterexamples to each ILP query, so
refinement scales with the number of mistakes to fix rather than the size of the
whole trace pool, and yields auditable edits to the existing rules instead of a
flat hypothesis that replaces them. ILP also runs off the critical path,
asynchronously alongside agent execution, so its cost does not affect runtime
enforcement.

\parh{Temporal and stateful properties.}
\tool evaluates predicates over individual events and their local context.
Properties that depend on history or state can already be encoded inside
predicates, and richer native temporal operators (e.g., LTL~\cite{pnueli1977temporal})
are a natural extension; because the CEGIS loop is agnostic to the constraint
language, such an extension reuses the same refinement algorithm.

\parh{Threats to validity.}
Our evaluation spans two domains, code execution and embodied agents, covering
diverse safety properties; other
settings such as web or database agents remain to be explored. The code-domain safe
traces mirror the RedCode~\cite{guo2024redcode} unsafe categories with benign
counterparts, deliberately stressing false positives, and labels rely on human
annotation. We mitigate both by drawing unsafe traces from an established
benchmark and supervising at trace granularity, the level at which operators
judge agent sessions.

%% file: related.tex
\section{Related Work}
\label{sec:related}

\parh{LLM Agent Safety.}
Existing defenses for LLM agents are authored statically and do not improve from
observed behavior: symbolic-rule DSLs~\cite{wang2025agentspec} and probabilistic
runtime monitors~\cite{wang2025probguard}, neural classifiers and LLM
judges~\cite{inan2023llama, bai2022constitutional, yuan2024r}, privilege,
access-control, and security-principle frameworks~\cite{ji2026taming,
shi2025progent, zhang2025llm}, multi-layered runtime
guardrails for foundation-model agents~\cite{shamsujjoha2025swiss}, and testing or red-teaming
harnesses~\cite{ruan2024identifying, perez2022red,wallace2019universal}. These defenses respond
to a rapidly expanding attack surface, from prompt
injection~\cite{liu2023prompt} to agent tool-protocol
threats~\cite{hou2025model}. To our knowledge, \tool is the first
\emph{safety-evolution} tool: it refines symbolic-rule guardrails from agent
traces, complementing these mechanisms, with the failures they expose (e.g.,
red-teaming traces) entering its loop as counterexamples.

\parh{Program Synthesis, Rule Learning, and Formal Methods.}
CEGIS~\cite{solar2006combinatorial,jha2010oracle} synthesizes programs from
specifications, and SyGuS~\cite{alur2013syntax} adds grammar-based search; on the
learning side, ILP systems such as Progol~\cite{muggleton1995inverse},
FOIL~\cite{quinlan1990learning}, Aleph~\cite{srinivasan2001aleph}, and
ILASP~\cite{law2020ilasp} learn logic programs from examples, while rule-based
classifiers like RIPPER~\cite{cohen1995fast} and CN2~\cite{clark1989cn2} learn
if-then rules from labeled data. \tool adapts CEGIS to safety rules with
domain-specific edit operations and uses ILASP to learn the discriminating
predicates that guide the loop, rather than synthesizing rules from scratch.
Complementarily, runtime verification~\cite{leucker2009brief} monitors execution
against temporal specifications (LTL~\cite{pnueli1977temporal},
MTL~\cite{koymans1990specifying}); \tool instead \emph{evolves} an existing set
of interpretable rules enforced at runtime, with temporal extensions left to
future work. More broadly, recent position papers argue that trustworthy LLM agents must
integrate formal methods with learning~\cite{zhang2024fusion}; \tool is a
concrete step, coupling inductive logic learning with counterexample-guided
synthesis to keep its guardrails adaptive yet interpretable.